# Distributed measurement of mode coupling in birefringent fibers with random polarization modes

Tianhua Xu*, Feng Tang, Wencai Jing, Hongxia Zhang, Dagong Jia, Xuemin Zhang, Ge Zhou, Yimo Zhang

National Education Ministry Key Laboratory of Optoelectronics Information and Technical Science, Tianjin University, Tianjin 300072, China

*Corresponding author: xutianhua0629@yahoo.com.cn

A scanning white light interferometer is developed to measure the distributed polarization coupling (DPC) in high birefringence polarization maintaining fibers (PMFs). Traditionally, this technique requests only one polarization mode to be excited or both polarization modes to be excited with equal intensity in the PMF. Thus, an accurate alignment of the polarization direction with the principal axis in PMF is strictly required, which is not facilely realized in practical measurement. This paper develops a method to measure the spatial distribution of polarization mode coupling with random modes excited using a white light Michelson interferometer. The influence of incident polarization extinction ratio (PER) on polarization coupling detection is evaluated theoretically and experimentally. It is also analyzed and validated in corresponding measurement that the sensitivity of the polarization coupling detection system can be improved more than 100 times with the rotation of the analyzer.

Keywords: white light inteferometry, birefringent fibers, polarization coupling, polarization extinction ratio, random polarization modes.

## 1. Introduction

Polarization maintaining fibers (PMFs) have played important roles in coherent communication, fiber-optic sensors, and integrated optics devices by providing high polarization conservation. However, when there are some internal or external perturbations, such as anisotropic Rayleigh scatting, transverse stress, micro-bending, twist and thermal fluctuations [1–3], optical power will couple from one polarization mode to the other one, which is called polarization mode coupling. This phenomenon degraded the capability of polarization preservation in PMFs [4, 5], and so the measurement of polarization mode coupling is significant in manufacturing of PMFs, and testing of PMF coils.

White light interferometry is widely used for measuring the spatial distribution of polarization coupling in PMFs, because this technique offers large dynamic range [6, 7], high accuracy and spatial resolution [8, 9], and is insensitive to optical power fluctuations [10]. It is usually noticed that only one polarization mode in the PMF should be excited or both polarization modes be excited with equal intensity when using this low coherence technique [7–10]. This needs precise alignment of the polarization direction with the principal axis of the PMF. It is not realized facilely in practical measurement. This paper developed a method to measure the spatial distributed polarization mode coupling using white light Michelson interferometry with random modes excited. The influence of the incident polarization extinction ratio (PER) on the measurement result was evaluated theoretically and experimentally. A polarization state adjusting mechanism is designed, and the analyzer can be oriented at any angle of the principle axes in PMF. By rotating the analyzer, the incident PER can be calculated, and the measurement sensitivity of the distributed polarization coupling (DPC) detection system can obviously be enhanced.

## 2. Model of polarization mode coupling in PMFs

Due to the geometrical effect of the core or the stress effect around the core, large modal birefringence of more than $10^{-4}$ was obtained in PMFs [3]. Within the undisturbed PMFs, there is minimal coupling between the perpendicular polarization modes, and the polarization coupling intensity is below –80dB everywhere along the fiber [4]. In the presence of external perturbation, the fiber can be modeled as a concatenation of three sections, as shown in Fig. 1.

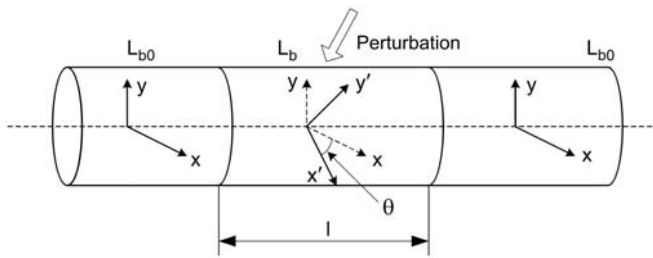

Fig. 1. Three sections model of PMF under perturbation.

Supposing the length of fiber in the undisturbed section is $l$. The other two fiber-sections without perturbation have the same beat length $L_{b0}$, and both birefringence axes in three sections are aligned. Under external perturbation, the beat length of the disturbed fiber section is converted into $L_b$, and the birefringence axes are rotated by an angle $\theta$ relative to the other two undisturbed sections. A fraction of the propagating optical power will be transferred from one polarization mode to the other mode, when a light beam passes through the disturbed region. The initial incident wave can be represented by Jones vector

$$\begin{bmatrix} E_x(0) \\ E_y(0) \end{bmatrix} = E_0 \begin{bmatrix} \cos\delta \\ \sin\delta \exp(i\Phi) \end{bmatrix} \quad (1)$$

where $E_0$ is the electric field magnitude of the incident beam, $\delta$ is the azimuth of the electric field vector with x-axis, and $\Phi$ is the phase difference between the two polarization components. Then the output electric field passing through the perturbed section can be expressed as the following,

$$\begin{bmatrix} E_x(l) \\ E_y(l) \end{bmatrix} = \begin{bmatrix} \cos\theta & -\sin\theta \\ \sin\theta & \cos\theta \end{bmatrix} \begin{bmatrix} \exp(ikn_o l) & 0 \\ 0 & \exp(ikn_e l) \end{bmatrix} \begin{bmatrix} \cos\theta & \sin\theta \\ -\sin\theta & \cos\theta \end{bmatrix} \begin{bmatrix} E_x(0) \\ E_y(0) \end{bmatrix} =$$

$$= \begin{bmatrix} E'_x + \exp(-ikn_o l)\sin\theta \cos\theta \left[1 - \exp\left(-\dfrac{i2\pi l}{L_b}\right)\right] E_x(0) \\ E'_y + \exp(-ikn_o l)\sin\theta \cos\theta \left[1 - \exp\left(-\dfrac{i2\pi l}{L_b}\right)\right] E_y(0) \end{bmatrix} \quad (2)$$

where $n_o$ and $n_e$ are the refractive indexes of the fast and slow axes in disturbed fiber section, respectively, and $E'_x$ is a component which has no relationship with $E_y(0)$, and the same is $E'_y$. Supposing the electric field coupling efficiency $C$ is the rate of magnitude transferring to the cross-polarization state. So the coupling efficiency between two orthogonal polarization modes is given by

$$C = C_{x \to y} = C_{y \to x} = \exp(-ikn_o l)\sin\theta \cos\theta \left[1 - \exp\left(-\dfrac{i2\pi l}{L_b}\right)\right] \quad (3)$$

Then the power coupling strength $h$ can be expressed as

$$h = h_{x \to y} = h_{y \to x} = |C|^2 = \sin^2(2\theta)\sin^2\left(\dfrac{\pi l}{L_b}\right) \quad (4)$$

As described in Eq. (4), the rotation of birefringence axes and the perturbation length are two key ingredients of the polarization mode coupling effect, and either ingredient can make the coupling intensity $h$ be zero. For example, if the perturbation length $l$ is the multiple of $L_b$, there is no mode coupling existing. Meanwhile, for a specific coupling point, whatever the excited polarization state, the coupling intensity from fast axis to slow axis or from slow to fast axis is the same.

## 3. Measurement principle of distributed mode coupling

The principle of distributed polarization coupling (DPC) detection system is expatiated in Fig. 2. Linear polarized broadband light is coupled into the PMF under test with only one polarization mode excited. When there is one polarization coupling point, a little fraction of light will be coupled into the other orthogonal mode. Because of the group modal birefringence $\Delta N_b$ of the fiber [11], two polarization modes travel at different group velocities. At the output end of the PMF, the optical path difference (OPD) $\Delta N_b \times l$ is generated between the two orthogonally polarized modes, where $l$ is the fiber length between the coupling point and the output fiber end; $l$ also represents the coupling point position. Using an analyzer, the two modes are projected to the same polarization direction. The OPD $\Delta N_b \times l$ is compensated by a scanning Michelson interferometer, and white light interferogram is recorded during the scanning process.

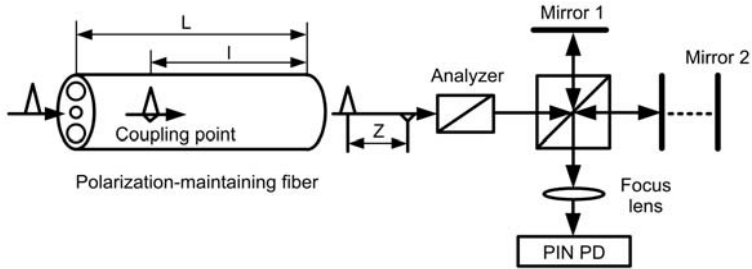

Fig. 2. Scheme of polarization coupling detection system.

When there is only one coupling point in the fiber, and the analyzer is oriented at an angle of 45°, the output interferogram from the Michelson interferometer is given by

$$I(d) = I_0 \left\{ |\gamma_0(d)| \cos(k_0 d) + \sqrt{h} \, |\gamma_0(\Delta N_b l - d)| \cos\left[k_0(\Delta n_b l - d)\right] \right\} + \\ + \sqrt{h} \, |\gamma_0(\Delta N_b l + d)| \cos\left[k_0(\Delta n_b l + d)\right] \qquad (5)$$

where $I_0$ is the DC component of interference, $\gamma_0(x)$ is the optical coherence function of the light source, $\Delta N_b$ and $\Delta n_b$ are group modal birefringence and phase modal birefringence of the fiber, respectively [11, 12], $d$ is the OPD of Michelson interferometer, $k_0 = 2\pi/\lambda_0$ is the wave number in free space, and $h$ is the power coupling strength. The recorded interference fringe contains three packets with the motion of scanning mirror as shown in Fig. 3. The maximal fringe $I_{main}$ appears when the Michelson interferometer is in equilibrium. The other two symmetrical packets $I_{coupling}$ appear when the OPD in PMF is compensated by the scanning interferometer.

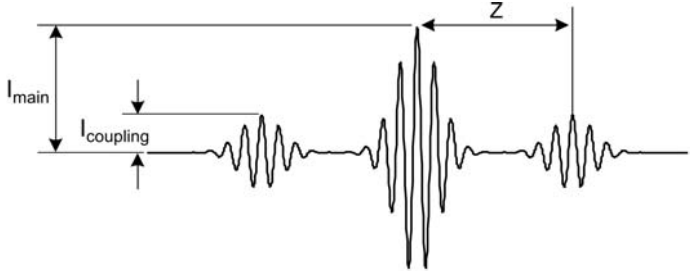

Fig. 3. Recorded interferogram with one coupling point existing.

The power coupling strength $h$ and coupling point position $l$ can be calculated with following equations:

$$h(\text{dB}) = 20 \log \frac{I_{coupling}}{I_{main}} \tag{6}$$

$$l = \frac{Z}{\Delta N_b} \tag{7}$$

where $Z$ is the peak distance between $I_{main}$ and $I_{coupling}$ in the recorded interferogram, as shown in Fig. 3.

## 4. Distributed measurement with random polarization modes excited

High spatial resolution and wide dynamic range can be achieved using traditional white light interferometry; coupling point location accuracy of ±1.5 cm and length resolution of better than 10 cm have been realized. However, it requests only one polarization mode in the PMF be excited or both polarization modes be excited equally. This requires precise alignment of the polarization direction with the principal axis in PMF. It is usually realized by polarization maintaining fusion splicer, which is not flexible and complicates system configurations. Here, a method for DPC detection with random polarization modes excited is developed, and the influence of input polarization extinction ration (PER) on distributed measurement of polarization mode coupling is analyzed.

### 4.1. Influence of the incident PER on the DPC detection

When both polarization modes are excited, the two orthogonal exciting modes can be expressed as

$$E_{x0} = A_{x0}\exp(i\Phi_{x0})$$
$$E_{y0} = A_{y0}\exp(i\Phi_{y0}) \qquad (8)$$

where $A_{x0}$ and $A_{y0}$, $\Phi_{x0}$ and $\Phi_{y0}$ are the initial magnitude and phase of the two orthogonal electric fields, respectively.

As shown in Fig. 4, when there exists one perturbation section, polarization coupling will occur in both axes of the PMF.

Generally, the perturbation fiber length is so short that the OPD between two polarization modes could be regarded to lie on the beat length of the undisturbed fiber. The output electric field from the fiber end can be calculated as,

$$E_x = \sqrt{1-h}\, A_{x0}\exp\left[i(\Phi_{x0} + k n_x l_{AB})\right] + \sqrt{h}\, A_{y0}\exp\left[i(\Phi_{y0} + k n_x l_{AB} - k\Delta n_b l_{AM})\right] \qquad (9)$$

$$E_y = \sqrt{1-h}\, A_{y0}\exp\left[i(\Phi_{y0} + k n_y l_{AB})\right] + \sqrt{h}\, A_{x0}\exp\left[i(\Phi_{x0} + k n_y l_{AB} + k\Delta n_b l_{AM})\right] \qquad (10)$$

where $\Delta N_b$ is the group refractive index difference between the two eigenmodes, $h$ is the power coupling strength, $l_{AB}$ is the total length of the disturbed fiber, $l_{AM}$ and $l_{BM}$ are the distances between incident mode and the two different orthogonal coupling modes, respectively, as described in Fig. 4. Interference will occur with OPD compensated by the scanning Michelson interferometer. The AC component of the interference packet is expressed as the following,

$$\begin{aligned}
I_{AC} = &\ \frac{1}{2}(1+h)\left(A_{x0}^2 + A_{y0}^2\right)\gamma(M_{OPD})\cos(k_0 M_{OPD}) + \\
&+ \sqrt{h}\, A_{x0} A_{y0}\gamma(\Delta N_b l_{AM} - M_{OPD})\cos\left[k_0(\Delta n_b l_{AM} - M_{OPD}) + \Phi_{x0} - \Phi_{y0}\right] + \\
&+ \frac{1}{2}\sqrt{h}\left(A_{x0}^2 + A_{y0}^2\right)\gamma(\Delta N_b l_{BM} - M_{OPD})\cos\left[k_0(\Delta n_b l_{BM} - M_{OPD})\right] + \\
&+ \frac{1}{2} A_{x0} A_{y0}\gamma(\Delta N_b l_{AB} - M_{OPD})\cos\left[k_0(\Delta n_b l_{AB} - M_{OPD}) + \Phi_{x0} - \Phi_{y0}\right] + \\
&+ \frac{1}{2} h A_{x0} A_{y0}\gamma\left[\Delta N_b(l_{BM} - l_{AM}) - M_{OPD}\right] \times \\
&\times \cos\left\{k_0\left[\Delta n_b(l_{BM} - l_{AM}) - M_{OPD}\right] + \Phi_{y0} - \Phi_{x0}\right\}
\end{aligned} \qquad (11)$$

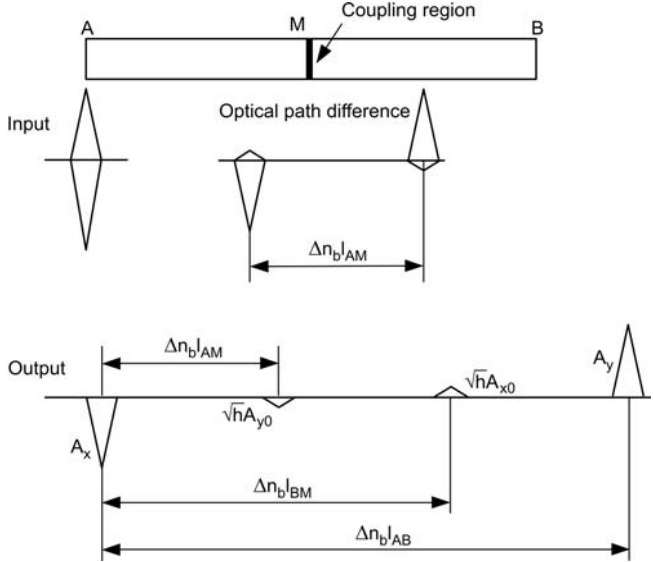

Fig. 4. Polarization coupling in PMF with both polarization modes excited.

where $M_{\text{OPD}}$ is the OPD introduced by the scanning Michelson interferometer, $k_0$ is the wave number in free space. It is found that the interference fringes contain five packets modulated by optical coherence function $\gamma(x)$. In the case of weak polarization coupling, the fifth interference package is so small as to be ignored, compared with others.

According to Eq. (6), the "coupling intensity" for the preceding four interference packages can be calculated as,

$$h_1 = \left[\frac{2\sqrt{h}\,A_{x0}A_{y0}}{(1+h)(A_{x0}^2+A_{y0}^2)}\right]^2 \approx \left[\frac{2\sqrt{h}\,\dfrac{A_{y0}}{A_{x0}}}{1+\left(\dfrac{A_{y0}}{A_{x0}}\right)^2}\right]^2 = \frac{4h \cdot \text{PER}}{(1+\text{PER})^2} \qquad (12)$$

$$h_2 = \left[\frac{\sqrt{h}}{1+h}\right]^2 \approx h \qquad (13)$$

$$h_3 = \frac{h_1}{4h} \approx \frac{\text{PER}}{(1+\text{PER})^2} \qquad (14)$$

$$h_4 = h_3 h^2 = \frac{h^2 \cdot \text{PER}}{(1+h)^2(1+\text{PER})^2} \qquad (15)$$

The first item $h_1$ corresponds to a fake interferential packet, which is related with both coupling intensity and input PER; the second item $h_2$ is the real coupling intensity corresponding to interested coupling point. The preceding two fringes are central symmetrical with middle of the fiber. The third one ($h_3$) is only related to the input PER, of which the packet position corresponds to the length of fiber. Since $h_3$ is only related to the input PER, the input PER can be calculated from $h_3$, as shown in the following equation:

$$\text{PER} = \frac{1 - 2h_3 + \sqrt{1 - 4h_3}}{2h_3} \tag{16}$$

If $\text{PER} \ll 1$, $h_3$ can be considered as the input PER. The fourth one could be ignored in most cases.

As mentioned above, when both polarization modes are excited randomly, a fake interference fringe package will be aroused, if the analyzer is oriented at 45° with the principal axis of PMF. This leads to a false coupling point judgment and the system dynamic range is decreased. The intensity decibel of the detected fake coupling point could be calculated from Eq. (12) approximately in the case of weak mode coupling,

$$h_1(\text{dB}) = 10\log h_1 \approx h(\text{dB}) - \left[\text{PER}(\text{dB}) - 6\right] \tag{17}$$

It could be found in Eq. (17) that the real coupling point with strength higher than (PER – 6) dB will lead to a fake coupling point, which will not occur when the coupling strength is lower than (PER – 6) dB. Hence, the dynamic range DR(dB) in the DPC detection system will be restricted according to

$$\text{DR(dB)} = \text{PER(dB)} - 6 \tag{18}$$

For example, if the incident PER is –25 dB, and the minimum detectable sensitivity of the system is –70 dB, then the system measurement dynamic range will be limited in 31 dB which is from –39 dB to –70 dB. Coupling point with coupling intensity larger than –39 dB will lead to a false weak coupling point.

**4.2. Improved DPC measurement with rotation of the analyzer**

In order to avoid the effect mentioned above, some improvements of the configuration illustrated in Fig. 2 have been carried out. The analyzer is mounted on a rotable stage driven by a step motor, which can be oriented at any angle to the axes of the PMF. The transmitted optical power will alternate like a sine wave with the analyzer rotating at an even speed, as shown in Fig. 5 and the following equation:

$$I_{\text{trans}} = (A_x \cos\varphi)^2 + (A_y \sin\varphi)^2 = A_x^2 \left[1 - (1 - \text{PER})\sin^2\varphi\right] \tag{19}$$

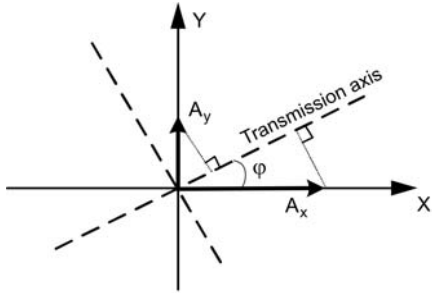

Fig. 5. Principle of analyzer adjusting.

The orientation of the analyzer can be calculated by detecting the transmitted optical power $I_{\text{trans}}$. The optical power reaches its maximum corresponding to the angle of 0°, and minimum corresponding to 90°. The middle value corresponds to the angle of 45°. Hence, the system PER can also be detected in the process of adjusting the analyzer. It will not affect the result whether the interference occurs.

When the analyzer orients at the angle of 0°, only $E_x$ component in Eq. (9) can enter into the Michelson interferometer. The AC component of the interference could be calculated as,

$$I_{\text{AC0}} = \left(A_{x0}^2 + h A_{y0}^2\right)\gamma(M_{\text{OPD}})\cos(k_0 M_{\text{OPD}}) +$$
$$+ \sqrt{h}\, A_{x0} A_{y0}\, \gamma(\Delta N_b l_{\text{AM}} - M_{\text{OPD}})\cos\left[k_0\left(\Delta n_b l_{\text{AM}} - M_{\text{OPD}}\right) + \Phi_{x0} - \Phi_{y0}\right] \tag{20}$$

Another interference packet and its coupling strength $h_5$ can be recorded,

$$h_5 = h \cdot \text{PER} \tag{21}$$

When the analyzer orients at 90°, the interference packet and its coupling intensity $h_6$ is obtained similarly,

$$I_{\text{AC90}} = \left(A_{y0}^2 + h A_{x0}^2\right)\gamma(M_{\text{OPD}})\cos(M_{\text{OPD}}) +$$
$$+ \sqrt{h}\, A_{x0} A_{y0}\, \gamma(\Delta N_b l_{\text{AM}} - M_{\text{OPD}})\cos\left[k\left(\Delta n_b l_{\text{AM}} - M_{\text{OPD}}\right) - \Phi_{x0} - \Phi_{y0}\right] \tag{22}$$

$$h_6 = \frac{h}{\text{PER}} \tag{23}$$

According to the preceding analysis, the system PER can be obtained by analyzer adjusting, and the coupling strength $h$ can be acquired by adjusting the analyzer at an angle between 0° and 90°. The system sensitivity can be improved to $h_m \cdot \text{PER}$, which is PER times higher than the traditional detection method.

## 5. Enhancement of coupling contrast ratio with the rotation of analyzer

The rotation of the analyzer not only can realize the measurement of polarization coupling with random modes excited, but also can improve the coupling interferential contrast ratio in the traditional DPC detection system. Equation (5) presents the previous relationship between coupling contrast ratio and coupling strength with the analyzer oriented at 45°. When the analyzer is oriented at a random angle, the coupling contrast ratio $\eta$ can be calculated as the following,

$$\eta^2 = \frac{I_{\text{coupling}}}{I_{\text{main}}} = \frac{h(1-h)\sin^2\varphi\cos^2\varphi}{\left[\cos^2\varphi - h\cos(2\varphi)\right]^2} \approx h\tan^2\varphi \qquad (24)$$

where $h \ll 1$ is the strength of the weak coupling point, and $\varphi$ is the angle between the analyzer and the principle axis in PMF in Fig. 5, $I_{\text{coupling}}$ and $I_{\text{main}}$ are the peak fringes of the coupling interferential packet and the main interferential packet, respectively, as shown in Fig. 3. The variation of coupling contrast ratio enhancement with the analyzer rotation angle $\varphi$ is illustrated in Fig. 6.

It is found that if the rotation angle of the analyzer is set at $\varphi = 75°$, the coupling contrast ratio $\eta$ of the output interferogram can be improved more than 10 times. Meanwhile, the sensitivity of the DPC detection system $h \approx \eta^2\cot^2\varphi$ can be enhanced more than 100 times. This method can be adopted for the detection of the weak mode coupling in birefringent fibers and waveguides, as well as sensitive distributed optical fiber sensors. Although the measurement sensitivity can be further enhanced with the rotation angle $\varphi$, it will induce the measurement error. The measurement error is dependent on the angle uncertainty $\Delta\varphi$. Their relationship can be described by

$$\text{Err} = \left| \frac{\cot^2(\varphi + \Delta\varphi) - \cot^2\varphi}{\cot^2(\varphi + \Delta\varphi)} \right| \tag{25}$$

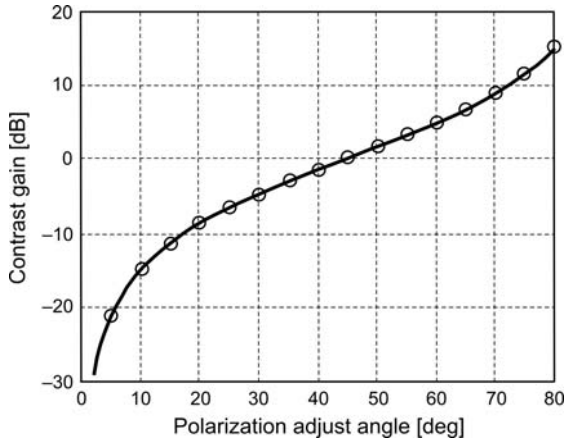

Fig. 6. Relationship between coupling contrast ratio enhancement and rotation angle.

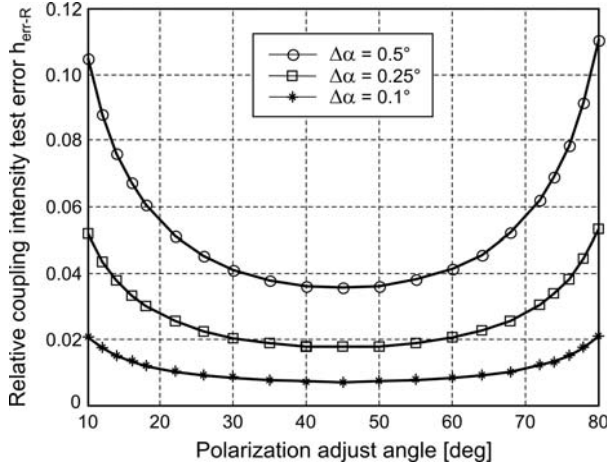

Fig. 7. Relationship between the relative measurement error and the rotation angle.

The relationship between the measurement error and the angle uncertainty is shown in Fig. 7. The minimum measurement error occurs at $\varphi = 45°$ for a given value of rotation angle uncertainty. This is the reason for previous works to choose $\varphi = 45°$ as the polarization eigenmodes rotation angle. The increment of the analyzer rotation angle $\varphi$ is adopted for the detection of weak polarization mode coupling. If the rotation angle is selected as $\varphi = 75°$, the measurement sensitivity can be improved more than 100 times. Meanwhile, the relative measurement error can still be kept within 5% for an angle uncertainty of 0.1°. Further increase in the rotation angle will cause a sharp-increase in the measurement error. So, a compromise should be considered between the measurement sensitivity and measurement error.

## 6. Experiments

A distributed polarization coupling test system is developed to validate the theory discussed above. A superluminescent diode (SLD) with a central wavelength of 1310 nm and spectrum width (FWHM) of 50 nm was used as the low-coherence light source. The scanning arm of the Michelson interferometer has a length of 300 mm. The instrument has the ability for testing 1 km PMF with spatial resolution better than 7 cm.

A polarization-maintaining patchcord with a length of 20 meters is used to carry out the experiment. The beat length of the PMF is about 2.1 mm. The recorded experiment results of the DPC detection are shown in Fig. 8. The interferogram is sampled with an interval of 400 nm in air.

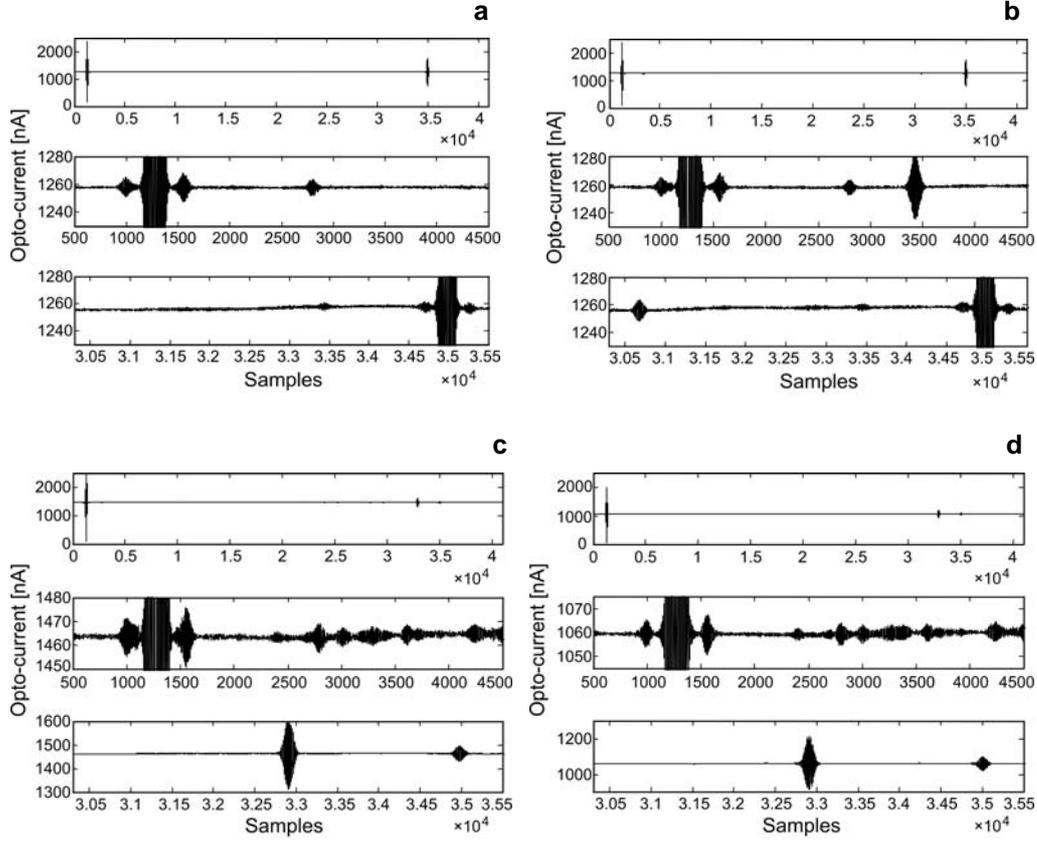

Fig. 8. Recorded interferograms of PMF with random modes excited; measurement at 45° without coupling point (**a**), measurement at 45° with one coupling point (**b**), measurement at 0° with one coupling point (**c**), measurement at 90° with one coupling point (**d**).

As shown in Fig. 8, part (**a**) is the interferogram for the patchcord with no external perturbation. In Figures 8**b**–8**d**, a clip is put onto the birefringent fiber to produce one coupling point with stable coupling strength and position. It is found that the experiment results achieve a good agreement with the theoretical analysis in Sec. 4. In Figure 8**b**, three interference packages are generated corresponding to one identical coupling point, and their positions are consistent with the description in Eq. (11). In Figures 8**c** and 8**d**, only one large interference package is produced by the coupling point, which is described in Eqs. (20) and (22). The sensitivity of this DPC detection system can reach –75 dB. It is notable that several coupling points, which are not found in Fig. 8**a**, are revealed in Figs. 8**c** and 8**d** very clearly. So it is also proved that the method developed in this paper can improve the sensitivity and dynamic range of the detection system. The improvement of coupling contrast ratio and measurement sensitivity with the rotation of analyzer is also validated in the experiments shown in Fig. 9, which achieves a good agreement with the foregoing theoretical analysis.

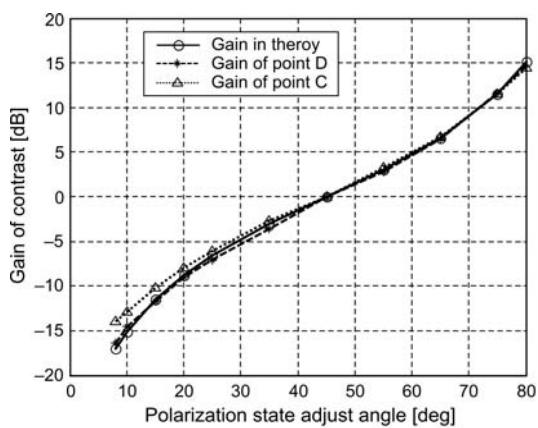

Fig. 9. Enhancement of contrast ratio with rotation of analyzer at two different points.

## 7. Conclusions

A white light interferometer has been designed and implemented to measure the distributed polarization mode coupling in high birefringence polarization maintaining fibers (HiBi-PMFs). The influence of the incident polarization extinction ratio (PER) on the measurement result was evaluated theo-retically and experimentally. A polarization state adjusting mechanism is designed to orient the analyzer at any angle to the principle axes of PMF. By alternating this angle between 0°, 90° and 45°, the incident PER can be detected and the measurement can be carried out with random polarization mode excited. The system sensitivity can also be improved more than 100 times with the rotation of the analyzer.

*Acknowledgements* – This work was supported by the National Natural Science Fund of China under contact No. 60577013, and the Teaching and Research Award Program for Outstanding Young Teachers in Higher Education Institutions of MOE, China.